\documentclass[a4paper]{article}

\usepackage{INTERSPEECH2022}
\usepackage{amsmath}
\usepackage{autobreak}

\usepackage{subfigure} 
\usepackage{cite}

\title{A CTC Triggered Siamese Network with Spatial-Temporal Dropout for Speech Recognition}
\name{Yingying Gao$^1$, Junlan Feng$^1$, Tianrui Wang$^{1,2}$, Chao Deng$^1$, Shilei Zhang$^1$}
\address{
  $^1$China Mobile Research Institute\\
  $^2$Institute of Information Science, Beijing Jiaotong University}
\email{\{gaoyingying, fengjunlan, dengchao, zhangshilei\}@chinamobile.com, wangtianrui.cd@outlook.com}

\begin{document}

\maketitle
\begin{abstract}
Siamese networks have shown effective results in unsupervised visual representation learning. These models are designed to learn an invariant representation of two augmentations for one input by maximizing their similarity. In this paper, we propose an effective Siamese network to improve the robustness of End-to-End automatic speech recognition (ASR). We introduce spatial-temporal dropout to support a more violent disturbance for Siamese-ASR framework. Besides, we also relax the similarity regularization to maximize the similarities of distributions on the frames that connectionist temporal classification (CTC) spikes occur rather than on all of them. The efficiency of the proposed architecture is evaluated on two benchmarks, AISHELL-1 and Librispeech, resulting in 7.13\% and 6.59\% relative character error rate (CER) and word error rate (WER) reductions respectively. Analysis shows that our proposed approach brings a better uniformity for the trained model and enlarges the CTC spikes obviously.
\end{abstract}
\noindent\textbf{Index Terms}: contrastive learning, Siamese network, spatial-temporal dropout

\section{Introduction}
The Siamese neural network was first proposed in \cite{tdnnSiam} for signature verification. It consists of two identical neural networks, which respectively takes as input one variant of the same entity and optimize model parameters to maximize the similarity between the outputs of the two neural networks.
Recently, various forms of Siamese networks \cite{simclr,moco,BYOL,swav,simsiam} have been invented to tackle multiple visual tasks such as face verification, tracking and others.
Siamese networks have also been introduced to speech representation learning \cite{drop_interspeech,phonetic,speaker1,speaker2,speaker3,speaker4,gender,emotion}. These tasks target on learning invariant acoustic features for classes such as phone\cite{drop_interspeech,phonetic}, speaker \cite{speaker1,speaker2,speaker3,speaker4}, gender \cite{gender} and emotion \cite{emotion}. 

Typically, inputs to Siamese networks are two randomly augmented views $x_1$ and $x_2$ from an image $x$ as shown in Figure 1. The model processes the two views by two encoder networks. The encoder could be a ResNet backbone, a convolution network or attention-based network \cite{simsiam}. The training objective is to maximize the similarity of the two output vectors on the left and right. Siamese networks have been widely proved as an effective design to model invariance, which is a key point of representation learning.

\begin{figure}[t]
  \centering
  \vspace{-0.4cm}
  \includegraphics[width=5cm]{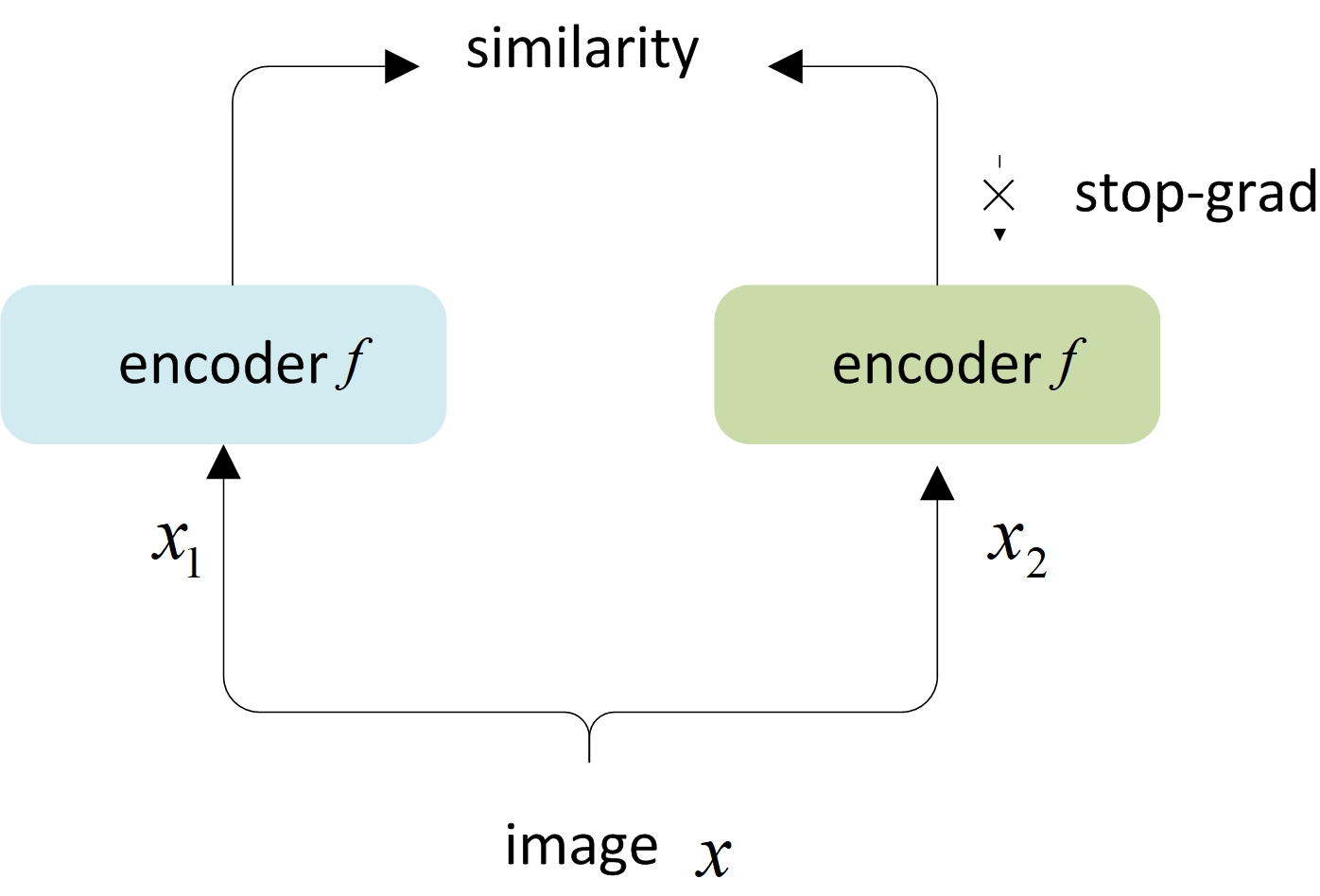}
  \caption{A typical architecture of Siamese network.}
  \label{fig0}
  \vspace{-0.5cm}
\end{figure}

Our main objective is to find an effective way to integrate Siamese network with current ASR models to improve ASR accuracy without paying a high cost, meaning without adding a substantial amount of model parameters. Inspired by the regularized dropout (R-drop) work reported in \cite{rdrop}, we choose to rely on the model itself instead of extra augmented data to increase the model robustness. 
However, unlike the positive results obtained in computer vision and NLP applications \cite{rdrop}\cite{simcse}, the achievements of Siamese networks in ASR are relatively insignificant or insufficient. In this paper, spatial-temporal dropout is introduced to construct ASR Siamese network, which can be performed in three modes: spatial, temporal, and both.
Instead of discarding discrete nodes randomly, it drops adjacent chunks of the representations in spatial and/or temporal domain, considering the spatial and/or temporal correlation of feature mappings to supply a better representation. Moreover, the similarity regularization in Siamese networks is mainly conducted on all the output representations, while in the CTC network for ASR, the network predicts the sequence of input tokens typically as a series of spikes and the most possible hypothesis is obtained directly from the CTC spikes. Therefore, in this work, the similarity regularization is implemented on the frames that CTC spikes occur rather than on all of them. Our proposal achieves 7.13\% and 6.59\% relative CER and WER reductions on two benchmarks, AISHELL-1 and Librispeech. Extensive experiments are designed to demonstrate the effects of dropout rates, data augmentation and data size.

\begin{figure*}[t]
 \centering
 \includegraphics{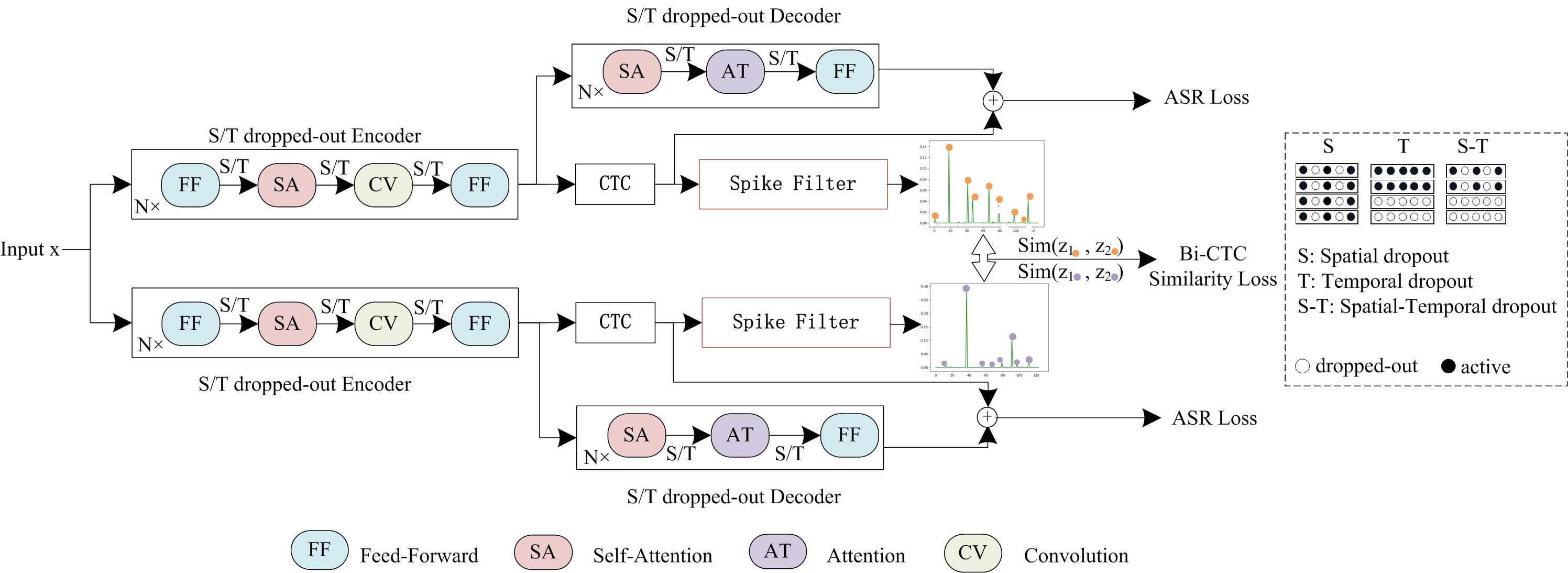}
 \caption{The proposed ASR-Siamese Model based on spatial-temporal droput and CTC triggered similarity loss.}
 \label{fig1}
\end{figure*}

\section{Our Proposal}
Our proposed Siamese model training architecture for ASR is shown in Figure~\ref{fig1}. 
The backbone comprises a typical ASR framework: a conformer encoder and a transformer decoder \cite{conformer}. Parameters of ASR models on two branches are shared. They differ only when dropout is applied to generate variations after each training example. Spatial and/or Temporal dropout is taken in either encoder or decoder or both of them to replace standard dropout. CTC spikes are detected on both the output distributions of the two encoders. And similarity loss is calculated on the peaked frames and their counterparts in the other submodel. The ASR loss is a combination of CTC and attention loss. And the overall loss is added with the similarity loss.

\subsection{Spatial-Temporal Dropout}
For each training example $x_i$ in a mini-batch, we create a copy of $x_i$ denoted as $x_{i+N}=x_i$, where N is the size of the mini-batch. With this setting, $x_i$ and $x_{N+i}$ obtains two different representations since the dropout operator drops different units in the two forward passes. For the same input data pair $(x_{i},y_{i})$, the output distribution of the two submodels are denoted as $z_{1}=P_{1}(y_{i}|x_{i})$ and $z_{2}=P_{2}(y_{i}|x_{i+N})$. Dropout here is used as a way to add variations to the output layer. One of the training losses is to minimize the similarity generated by dropout. 

As mentioned above, standard dropout somehow fails to supply significant disturbances to augment the input. Meanwhile, we can not achieve performance improvement by enlarging the dropout rate directly. Instead, we try to dropout a whole vector or several vectors in the feature mapping matrix, which provides considerable benefits. We find the implementation is homologous with spatial dropout \cite{spatial} which was proposed to enhance spatial correlation of convolution feature maps. In this work, we extent the convolution spatial dropout to a 2-dimentional space, since the input feature in ASR is a spatial-temporal spectrum. As shown in Figure~\ref{fig1}, vertical axis refers to temporal domain and the horizontal axis denotes spatial domain. All the pixels in one column or one row are either all zero (white) or all active (black). Additionally, we spread the spatial-temporal dropout to each module in encoder and decoder other than only in the convolution module.  

\subsection{CTC Triggered Similarity Regularization}
\subsubsection{Similarity Metric Selection}
Typically, a regularization term is added to the pure ASR loss. Here we employ a similarity-based regularization loss as baseline. 
Cosine similarity $S_{cs}$ and K-L divergence $S_{KL}$ are the two popular similarity regularization manners utilized in Siamese network. They are respectively defined as:
\begin{equation}
 S_{cs}(z_{1},z_{2})=\frac{z_{1}}{\left \| z_{1} \right \|^{2}}\cdot \frac{z_{2}}{\left \| z_{2} \right \|^{2}}=\frac{\sum_{j=1}^{D}z_{1,j}z_{2,j}}{\sum_{j=1}^{D}z_{1,j}^{2}\sum_{j=1}^{D}z_{2,j}^{2}} 
  \label{eq4}
\end{equation}
\begin{equation}
 S_{KL}(z_{1}\|z_{2})=z_{1}\log\frac{z_{1}}{z_{2}}=\sum_{j=1}^{D}z_{1,j}\log\frac{z_{1,j}}{z_{2,j}} 
  \label{eq6}
\end{equation}
\vspace{-0.4cm}
where $D$ is the dimension of encoder output. 

We find that the model with cosine similarity loss converges faster than the one regularized by K-L divergence. Moreover, cosine similarity loss is able to prevent collapsing even without stop gradient, while the K-L divergence fails. We analyze their gradients in order to understand the possible reason behind the the above observation.
For any dimension $j$, the gradient on $z_{1,j}$ from $S_{KL}$ is:
\begin{equation}
\frac{\partial S_{KL}}{\partial z_{1,j}}=\log z_{1,j}+1-\log z_{2,j}-\frac{z_{1,j}}{z_{2,j}}
\label{eq8}
\end{equation}
If $z_{1,j}=z_{2,j}$, then $\frac{\partial S_{KL}}{\partial z_{1,j}}=0$, meaning if the two values on the same dimension are equal, the gradient on that dimension will disappear.
And the gradient on $z_{1,j}$ from $S_{cs}$ is
\begin{equation}
\begin{split}
& \frac{\partial S_{cs}}{\partial z_{1,j}}=  \\
& \frac{z_{2,j}\sqrt{\sum_{k=1}^{D}{z_{1,k}^{2}}}-z_{1,j}\sum_{k=1}^{D}(z_{1,k}z_{2,k})/\sqrt{\sum_{k=1}^{D}{z_{1,k}^{2}}}}{\sum_{k=1}^{D}{z_{1,k}^{2}}*\sqrt{\sum_{k=1}^{D}{z_{2,k}^{2}}}} \\
\label{eq9}
\end{split}
\end{equation}
implying that the gradient from $S_{cs}$ will vanish unless the values of $z_{1,k}$ and $z_{2,k}$ on all the $D$ dimensions are equal, and this is harder to achieve than $S_{KL}$. 
In our experiments, we select $S_{cs}$ as the similarity metric.

\subsubsection{CTC Triggered Cosine Similarity}
The standard similarity evaluation in Siamese network is carried out on all the frames of the two feature mappings. However, in the CTC network for ASR, instead of a frame-wise alignment, the CTC model predicts the sequence of input tokens typically as a series of spikes, and the most probable result is obtained directly from the CTC spikes. Therefore, the similarity comparison between the two outputs on each frame is regarded as unnecessary since most of them denote blanks. Only the frames at CTC spikes are considered in our cosine similarity. Since the output probabilities of the two branches might be different, the spikes on both of them are detected. Cosine similarities are calculated on these frames and their counterparts from the other branch.
A CTC spike filter based on matrix is built to decrease time consumption. The implementation is shown in Algorithm 1. $z_{1sp}$ refers to the outputs of one encoder at spikes, and $z_{2co}$ stands for the corresponding outputs of the other encoder at the same frames.

\begin{table}[ht]
\centering
  \label{tab2}
\begin{tabular}[H]{l}
\hline
\textbf{Algorithm 1:} The implementation of spike filter \\
\hline
\textbf{Input:} $P$: the output probabilities of CTC \\
$z_1$, $z_2$: the outputs of encoder 1 and encoder 2 \\
$sp_1$, $sp_2$: the spike indexes for encoder 1 and encoder 2   \\
\textbf{Initialise:} left shift of $P$: $P_l \gets zeros(size\,of\, P$) \\
right shift of $P$: $P_r \gets zeros(size\,of\, P$) \\
\textbf{Step1: left and right shift for $P$} \\
$P_l[:,1:] \gets P[:,:-1]$ \quad $P_r[:,:-1] \gets P[:,1:]$  \\
\textbf{Step2: search spike frames} \\
\# set spike framses as 1, set others as 0  \\
$sp \gets -(sign((sign((P_l-P)*(P-P_r))+0.1))-1)/2$  \\
$sp_1 \gets sp[:batchsize]$ \quad $sp_2 \gets sp[batchsize:]$  \\
\textbf{Step3: filter the outputs of encoder} \\
$z_{1sp} \gets z_1*sp_1$ \quad $z_{2co} \gets z_2*sp_1$  \\ 
$z_{2sp} \gets z_2*sp_2$ \quad $z_{1co} \gets z_1*sp_2$  \\
\hline
\end{tabular}
\end{table}

The similarity loss regularization $L_{sim}$ based on $S_{cs}$ is modified to:
\begin{equation}
L_{sim}=-\frac{1}{2}(S_{cs}(z_{1sp},z_{2co})+S_{cs}(z_{1co},z_{2sp}))
\end{equation}

The entire model is trained to minimize the similarity loss $L_{sim}$ between the output probabilities of two sides and the typical ASR loss $L_{asr}$, which is a weighted average of CTC loss $L_{ctc}$ and attention loss $L_{att}$ \cite{joint}:
\begin{equation}
 L_{asr}=\alpha*L_{ctc}+(1-\alpha)*L_{att} 
\end{equation}

The final training loss $L$ is:
\begin{equation}
 L= L_{asr} +\lambda *L_{sim}
\end{equation}
Both the $L_{asr}$ from the two branches are considered since the input $x$ are forwarded to both of them. During the inference stage, only one branch is remained. 

\section{Experiments}
First, we will show the performance comparison of the proposed Siamese model against the other original Siamese networks in ASR task. Then we investigate the effects of dropout rate, data augmentation and data size. Finally, we analyze the benefit of the proposal from the perspective of alignment-uniformity and CTC spike, in which the former constitutes the two key properties related to contrastive learning, and the later is directly related with the prediction result in ASR.

\textbf{Dataset and Experimental Setup:} Our experiments are carried out on two open-source benchmarks AISHELL-1 \cite{aishell} and Librispeech \cite{libri}. AISHELL-1 contains 178 hours of Mandarin speech and Librispeech is a corpus of 960 hours of English speech. We select Wenet \cite{wenet} as the ASR training toolkit. The basic ASR network consists of 12 conformer blocks as the encoder followed by 6 transformer blocks as the decoder. The combination ratio $\alpha$ for CTC loss is 0.3 and the weight $\lambda$ for $L_{sim}$ is 0.1. Four decoding modes are supported in Wenet. The results decoded by ctc greedy search algorithm (CTC) and attention rescoring (ATT-res) are reported, in which the speed of the former is faster while the performance of the latter is better. 

\subsection{Network Comparison}
First, we compare various Siamese networks with our proposed method. The baseline is the ASR model trained by non-Siamese end-to-end model. Data augmentation is turned off to focus on Siamese network itself. CTC spikes can only be obtained from the output of encoder, thus the similarity loss is calculated on the outputs of encoders. The results shown in Table~\ref{tab1} display that the primitive Siamese networks obtains slight improvements on ASR task. DropC denotes the Siamese network built upon dropout operation and cosine similarity. The bi-directional CTC triggered similarity (BiCTC-DropC) shows a 3.40\% relative improvement on CER over DropC and performs slightly better than the network detecting CTC spikes on one branch (CTC-DropC). Temporal dropout (T-DropC) presents a 9.01\% relative gain on CER compared with DropC, which is superior to spatial dropout (S-DropC) and the combination of them (S-T-DropC). At last, the integration of temporal dropout and bi-directional CTC similarity obtains the optimal result, improving the ASR performance by 12.94\% and 11.52\% relative CER reductions compared to the non-Siamese baseline and the DropC network.


\begin{table}[ht]
\centering
\caption{The character error rates (CERs) and relative reductions (CERRs) compared to baseline on AISHELL-1 test set via the proposal and other Siamese networks.}
  \label{tab1}
\begin{tabular}{lcccc}
\hline
            & CTC & CERR & ATT-res  & CERR  \\
\hline
Baseline & 6.88 & & 6.56 & \\
\hline
SimCLR$^{[2]}$   & 13.77 &  & 12.01 &    \\
MoCo$^{[3]}$     & 9.28  &   & 8.35  &   \\
BYOL$^{[4]}$     & 8.14  &  & 7.65   &   \\
SimSiam$^{[6]}$  & 6.79  & 1.31  & 6.55  & 0.15 \\
R-Drop$^{[15]}$   & 6.69 & 2.76 & 6.31 & 3.81 \\
\hline 
DropC    & 6.77    & 1.60 & 6.31  & 3.81 \\
CTC-DropC   & 6.55   & 4.80 & 6.20 & 5.49 \\
\textbf{BiCTC-DropC} & \textbf{6.54}   & \textbf{4.94}  & \textbf{6.18} & \textbf{5.79}  \\
S-DropC & 6.29   &  8.58 & 6.02 & 8.23 \\
\textbf{T-DropC}  & \textbf{6.16}   & \textbf{10.47} & \textbf{5.89} & \textbf{10.21} \\
S-T-DropC & 6.22    & 9.59 &  5.95 &  9.30\\
\textbf{BiCTC-T-DropC}  & \textbf{5.99} & \textbf{12.94} & \textbf{5.75} & \textbf{12.35}\\
\hline
\end{tabular}
\end{table}

\subsection{Dropout Rate}
Besides the dropout mode, we investigate the impact of dropout rate on the construction of Siamese networks based on dropout. The first two rows in Table~\ref{tab2} demonstrates that enlarging the dropout rate fails to get positive results. The best setup of dropout rate for spatial and temporal dropout is 0.2 in the current comparison. Combining the two dropout (S-T-DropC) modes shows an improvement when dropout rate is smaller, but its best result is still inferior to T-DropC.

\begin{table}[ht]
\centering
\caption{The CERs (CTC greedy search) on AISHELL-1 test set via different dropout rates.}
  \label{tab2}
\begin{tabular}{lccc}
\hline
            & 0.1 & 0.2 & 0.3   \\
\hline
Baseline(Drop)  & 6.88  & 6.89  &  6.94 \\
DropC           &  6.77  &  6.88 &  7.16 \\
S-DropC     & 6.39    & \textbf{6.29}  &  6.79  \\
T-DropC    & 6.56    & \textbf{6.16}  &  6.38   \\
S-T-DropC   & \textbf{6.34}  & \textbf{6.22} & 7.55 \\
\hline
\end{tabular}
\end{table}

\subsection{Data Augmentation}
SpecAug \cite{specaug} and speed perturbation \cite{speed} are two common augmentation methods for ASR. In this test, we investigate the interactions between the proposed Siamese network and data augmentations. The results in Table~\ref{tab3} manifest that consistent improvement is produced by CTC triggered similarity while the impact of spatial-temporal dropout is weakened when data augmentation is added. Therefore, instead of adopting spatial-temporal dropout in the entire network as in previous experiments, we exploit it in some specific positions for a precise optimization. The results listed in Table~\ref{tab4} present that introducing spatial dropout and temporal dropout in both encoder and decoder (S-Drop and T-Drop) is detrimental to the performance. Conducting it in the convolution module in encoder (since it is originally proposed for CNN) (ConvS-Drop) is advantageous. EncS-Drop replaces dropout by spatial dropout in the overall encoder and exceeds the performance of ConvS-Drop. Finally, BiCTC-EncS-DropC reduces the CER by 7.13\% relative compared with the result of non-Siamese network.

\begin{table}[ht]
\centering
\caption{The recognition results (CERs) of the proposal on AISHELL-1 test set with data augmentations.}
  \label{tab3}
\begin{tabular}{lcccc}
\hline
            & CTC & CERR & ATT-res & CERR   \\
\hline
Baseline & 5.89 & & 5.25 & \\
\hline
DropC    & 5.67 & 3.74 & 5.19 &  1.14 \\
BiCTC-DropC & 5.53 & 6.11 & 4.99 & 4.95 \\
\hline
T-DropC   & 5.55  & 5.77 & 5.02 &  4.38  \\
S-DropC     & 5.54  & 5.94  & 4.99  &  4.95 \\
BiCTC-S-DropC     & 5.53  &  6.11 & 5.02   &  4.38 \\
\hline 
EncS-DropC   & 5.51  & 6.45 & 5.04 &  4.00 \\
\textbf{BiCTC-EncS-DropC}  & \textbf{5.47} &  \textbf{7.13} & \textbf{5.01} &  \textbf{4.57}\\
\hline
\end{tabular}
\end{table}
\vspace{-0.4cm}

\begin{table}[ht]
\centering
\caption{The recognition results (CERs) of different dropout positions on AISHELL-1 test set without Siamese architectures.}
  \label{tab4}
\begin{tabular}{lcc}
\hline
            & CTC & ATT-res    \\
\hline
Baseline & 5.89  & 5.25 \\
\hline
S-Drop   & 6.06   & 5.56    \\
T-Drop  & 6.10  &  5.53  \\
ConvS-Drop    & 5.96     & 5.30     \\
\textbf{EncS-Drop}   & \textbf{5.86}    & \textbf{5.32}   \\
EncT-Drop   & 5.96  & 5.36 \\
\hline
\end{tabular}
\end{table}
\vspace{-0.4cm}

\subsection{Data Size}
In this study, we evaluate the performance of the proposed network on Librispeech to investigate the effect of larger data size. Data augmentation is conducted in this experiment and spatial-temporal dropout is executed in the entire network. Table~\ref{tab5} presents that the impact of temporal dropout (T-Drop) is more obvious than on small corpus. The proposed network (BiCTC-T-DropC) generates 5.85\% relative WER reduction on test clean set and 6.59\% on test other set.

\begin{table}[ht]
\centering
\vspace{-0.4cm}
\caption{The attention rescoring WERs of the proposal on Librispeech test sets.}
  \label{tab5}
\begin{tabular}{lcc}
\hline
            & Test clean & Test other  \\
\hline
Baseline & 3.42 &  9.25 \\
\hline
S-Drop    & 3.47&   9.28  \\
\textbf{T-Drop} & \textbf{3.27} &\textbf{8.96} \\
S-T-Drop & 3.42 &   9.05 \\
\hline
DropC   &  3.25 &  9.00 \\
BiCTC-DropC    &  3.27 &  8.89   \\
\textbf{BiCTC-T-DropC}     & \textbf{3.22} &  \textbf{8.64}  \\
\hline 
\end{tabular}
\end{table}
\vspace{-0.4cm}

\subsection{Analysis}
\subsubsection{Alignment-Uniformity}
Alignment and Uniformity are the two key properties related to contrastive learning. Alignment requires that two samples forming a positive pair should be mapped to nearby features, and uniformity means that feature vectors distribute uniformly on the unit hypersphere and preserve as much information of the data as possible \cite{ali-uni}. For both alignment and uniformity, lower numbers are better. The alignment of the basic Siamese architecture (DropC) on both the two corpus are already zero so the effect of the proposal on alignment is not revealed. Figure~\ref{fig2} (a) shows that the CTC regularization brings a better uniformity for the trained model and the combination with spatial-temporal dropout makes the value even lower.
\subsubsection{CTC Spikes}
\begin{figure}[ht]
\centering
\vspace{-0.3cm}
\subfigure[Uniformity]{
\begin{minipage}[b]{0.22\textwidth}
\includegraphics[width=1\textwidth]{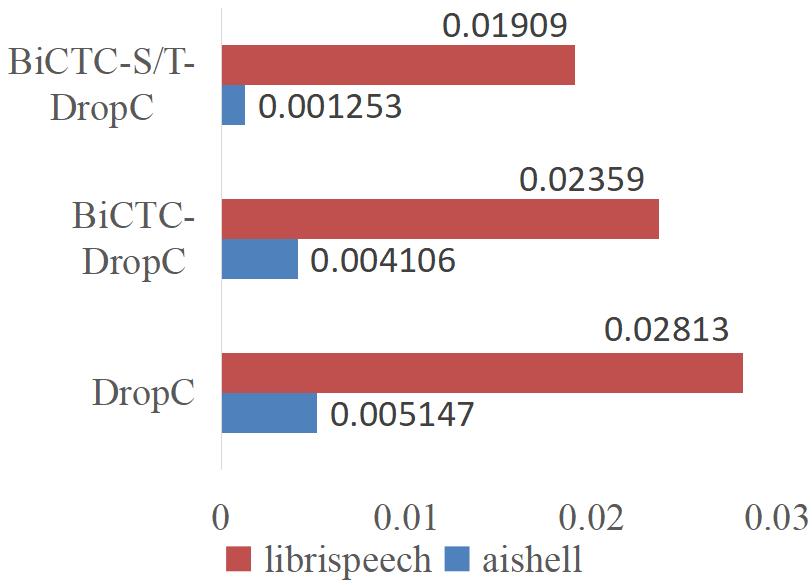} 
\end{minipage}
}
\subfigure[CTC spikes]{
\begin{minipage}[b]{0.22\textwidth}
\includegraphics[width=1\textwidth]{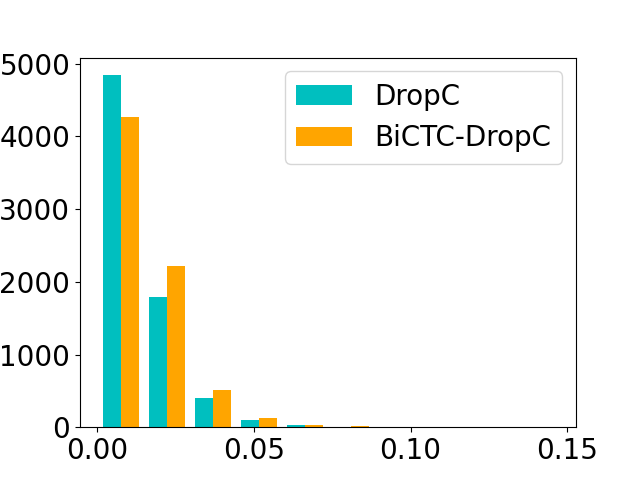} 
\end{minipage}
}
\vspace{-0.4cm}
\caption{The uniformity and CTC spike distribution of the proposal and baseline.}
\label{fig2}
\end{figure}

\begin{figure}[ht]
\centering
\vspace{-0.3cm}
\subfigure[DropC]{
\begin{minipage}[b]{0.22\textwidth}
\includegraphics[width=1\textwidth]{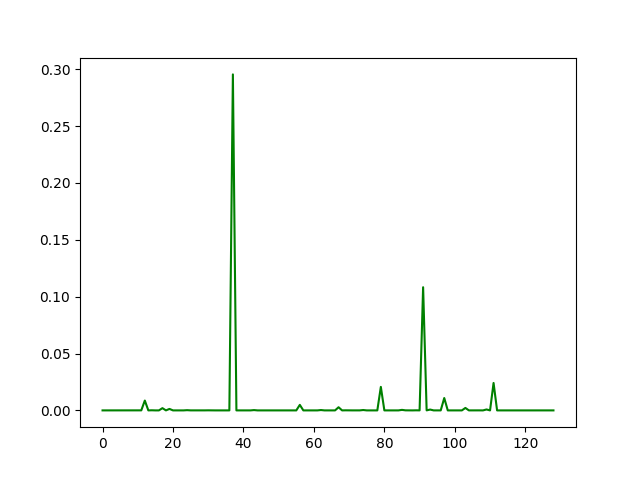} 
\end{minipage}
}
\subfigure[BiCTC-DropC]{
\begin{minipage}[b]{0.22\textwidth}
\includegraphics[width=1\textwidth]{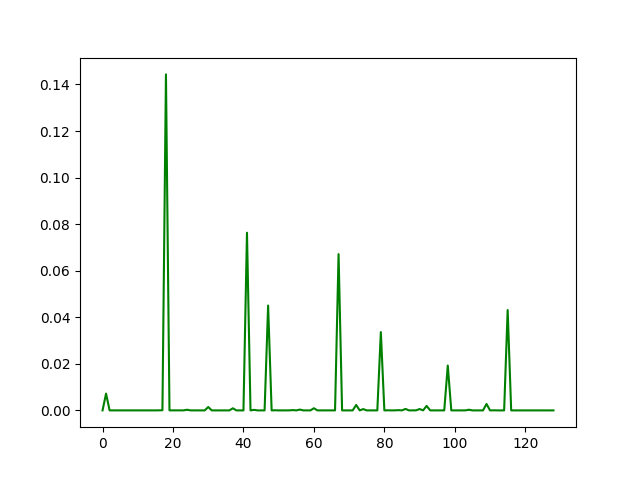} 
\end{minipage}
}
\vspace{-0.4cm}
\caption{The negative log CTC probabilities from DropC and BiCTC-DropC for one input.}
\label{fig3}
\end{figure}

In the CTC-based ASR model, the input sequence is predicted as a series of spikes and the most possible labelling is read directly from them. Therefore the probabilities of CTC spikes are related with the model performance and robustness. Through our observation, we find that the CTC similarity makes the CTC spikes more obvious (one sample is presented in Figure~\ref{fig3}). And the histogram of them from DropC and BiCTC-DropC (in Figure~\ref{fig2} (b)) also demonstrates that more spikes of BiCTC-DropC are distributed in the regions with larger possibilities.

\vspace{-0.4cm}

\vspace{-0.4cm}

\section{Conclusions}
In this paper, we propose a  workable Siamese network for ASR. 
We utilize spatial-temporal dropout to replace standard dropout, which is able to support better augmentations for the same input. Moreover, only the similarities of distributions on CTC spikes are regularized, which enlarges the probabilities of CTC spikes. Both spatial-temporal dropout and CTC triggered regularization are beneficial for the uniformity of the trained model. We will integrate the proposal with other Siamese networks and test the impacts on more and larger dataset in the future.

\bibliographystyle{IEEEtran}

\bibliography{myref}


\end{document}